\documentclass[10pt,twocolumn]{article}
\usepackage[margin=0.75in]{geometry}
\usepackage{times}
\usepackage{microtype}
\usepackage{amsmath,amssymb}
\usepackage{booktabs}
\usepackage{tabularx}
\usepackage{array}
\usepackage{multirow}
\usepackage{makecell}
\usepackage{enumitem}
\usepackage{graphicx}
\usepackage{xcolor}
\usepackage{tikz}
\usepackage{float}
\usetikzlibrary{arrows.meta,positioning,fit,shapes.geometric,calc}
\usepackage[hyphens]{url}
\usepackage[round,sort&compress]{natbib}
\usepackage[colorlinks=true,linkcolor=blue,citecolor=blue,urlcolor=blue]{hyperref}

\newcommand{\llm}{LLM}

\newcommand{\rag}{RAG}

\newcolumntype{Y}{>{\raggedright\arraybackslash}X}
\newcolumntype{L}[1]{>{\raggedright\arraybackslash}p{#1}}

\title{A Lifecycle and Application-Stack Survey of Large Language Model Vulnerabilities:\\ Attacks, Risks, Defenses, and Open Problems}

\author{
Seyed Bagher Hashemi Natanzi,\quad Bo Tang\\
\small Department of Electrical and Computer Engineering, Worcester Polytechnic Institute, Worcester, MA, USA\\
\small \texttt{\{snatanzi,btang1\}@wpi.edu}
}

\date{}

\begin{document}
\maketitle

\begin{abstract}
Large language models are no longer only text generators. They are increasingly embedded in retrieval pipelines, enterprise assistants, coding environments, robotic systems, security-operation workflows, and autonomous agents that can read private data, call tools, write files, execute code, and act across organizational boundaries. This shift changes the security problem: risks do not arise from the model weights alone, but from the full lifecycle and application stack through which data, prompts, model outputs, tools, memories, and user authority interact. This paper systematizes the literature on vulnerabilities in large language model systems through a lifecycle and application-stack lens. We organize attacks across eight stages: data collection, pretraining, post-training alignment, model packaging and supply chain, retrieval and memory, prompting and inference, tool/agent execution, and deployment/maintenance. For each stage, we analyze attacker capabilities, affected security objectives, representative attacks, practical risks, evaluation practices, and defenses. We further map LLM-specific vulnerabilities to confidentiality, integrity, availability, safety, privacy, fairness, accountability, and agency-control objectives. Unlike taxonomies that list isolated attack names, the proposed systematization emphasizes where trust boundaries fail, how untrusted data becomes executable instruction, how delegated authority amplifies model errors, and why point defenses rarely compose. We close with a research agenda for secure LLM systems, including compositional security, provenance-aware retrieval, tool-call containment, long-horizon agent evaluation, privacy-preserving adaptation, realistic red teaming, and deployment-grade incident response.
\end{abstract}


\section{Introduction}

Transformer-based language models have become a general-purpose substrate for natural-language interfaces, information retrieval, code generation, planning, and multimodal reasoning \citep{vaswani2017attention,devlin2019bert,brown2020language,openai2023gpt4,geminiteam2023gemini}. Their adoption has also shifted from model-centric research demonstrations to application stacks that combine model APIs, retrieval systems, user profiles, persistent memories, external tools, plug-ins, workflow orchestrators, and autonomous agents \citep{lewis2020retrieval,yao2022react,schick2023toolformer,qin2023toolllm}. This deployment shift changes the nature of security. A model that only emits text may cause unsafe content or misinformation; an application that lets the same model access files, email, calendars, databases, cloud APIs, browsers, terminals, or robots can convert text-generation failures into external side effects.

The resulting risk landscape is broad. Training data can contain private information, copyrighted content, low-quality sources, malicious triggers, or biased patterns \citep{bender2021dangers,bommasani2021opportunities,weidinger2021ethical}. Pretrained models can memorize rare sequences, reveal training examples, or expose privacy information through extraction and inference attacks \citep{carlini2019secret,carlini2021extracting,shokri2017membership}. Alignment and instruction tuning can reduce many harms but can also be attacked, bypassed, or weakened by fine-tuning \citep{ouyang2022training,bai2022constitutional,rafailov2023direct,qi2023fine}. Prompting interfaces are vulnerable to adversarial suffixes, jailbreaks, prompt leakage, direct prompt injection, and indirect prompt injection through untrusted external content \citep{perez2022ignore,greshake2023not,liu2023prompt,zou2023universal,wei2023jailbroken}. Retrieval-augmented generation (RAG) introduces document ingestion, embedding, vector search, provenance, and stale-memory risks \citep{lewis2020retrieval,karpukhin2020dense,xiong2024ragsecurity,zhao2024ragpoisoning}. Tool-using agents add delegated authority, state, persistent memory, multi-step planning, and protocol-level attack surfaces \citep{yao2022react,ferrag2025agentthreats,ling2026secureagents,chu2026systematicagents,zhao2026clawguard}.

Existing surveys have made important progress, but the field remains fragmented. Some surveys organize risks by attack technique, such as prompt injection, poisoning, jailbreaking, or privacy leakage \citep{yao2024llmsecurity,xu2025surveyattacks}. Others focus on LLMs in general, alignment, trustworthiness, or agent-specific security \citep{naveed2023comprehensive,liu2023trustworthy,derner2024security,ling2026secureagents,chu2026systematicagents}. Industry frameworks such as OWASP's LLM Top 10 and NIST's adversarial machine-learning taxonomy provide operational vocabulary and mitigation categories \citep{owasp2025llm,vassilev2025adversarial}. However, developers and researchers still need a unifying systems view that answers not only \emph{what} the attack is called, but \emph{where} it arises in the lifecycle, \emph{which} trust boundary failed, \emph{what} authority the model had, and \emph{how} defenses should compose across the stack.

We argue that LLM vulnerabilities are best understood as failures of information flow, provenance, privilege, and state across a lifecycle and application stack. The same surface string can be harmless in a sandboxed chatbot but dangerous in an agent with database write privileges. Similarly, the same poisoning attempt has different implications when it appears in pretraining data, an instruction-tuning set, a RAG corpus, a tool description, or a persistent memory. A systematization-of-knowledge survey should therefore connect attacks to lifecycle stage, security objective, attacker capability, and defense layer. In summary, this paper makes four contributions.
\begin{enumerate}[leftmargin=*]
    \item We propose a lifecycle and application-stack taxonomy for LLM vulnerabilities, covering data collection, pretraining, post-training alignment, model packaging and supply chain, RAG/memory, prompting and inference, tool/agent execution, and deployment/maintenance.
    \item We map attacks to security objectives beyond the traditional confidentiality--integrity--availability triad, adding safety, privacy, fairness, accountability, and agency control as first-class objectives for LLM applications.
    \item We synthesize defenses into a defense-in-depth architecture that separates deterministic controls, model-level robustness, monitoring/red teaming, privacy controls, and governance processes.
    \item We identify open problems that remain under-specified in current literature, including compositional prompt-injection defenses, provenance-preserving RAG, secure tool delegation, stateful/multi-agent evaluation, supply-chain integrity for model artifacts, and incident response for adaptive LLM systems.
\end{enumerate}

\section{Scope, Method, and Terminology}

\subsection{Scope}

We use \emph{large language model system} to refer to a deployed system that includes one or more language or multimodal foundation models, prompts, safety policies, retrieval components, context windows, memory stores, tool-call interfaces, plug-ins, monitoring components, and human operators. This scope is broader than the base model. It includes model weights, but also the application stack in which the model is embedded.

We focus on vulnerabilities that are specific to, amplified by, or operationally transformed by LLMs. General web, cloud, identity, and software-security vulnerabilities are included only when LLM integration changes the attack path or impact. For instance, SQL injection is a classical vulnerability; it becomes part of this survey when an LLM generates SQL from natural language or when untrusted database content becomes part of an LLM prompt. We also include multimodal and agentic systems when the LLM acts as the central planner or interface.

\subsection{Review protocol and coding schema}

The goal is systematization rather than exhaustive bibliometrics. We used seed papers from LLM security, privacy, adversarial ML, RAG, and agent-security literature; expanded by citation chasing; and cross-checked against practitioner taxonomies including OWASP LLM Top 10, MITRE ATLAS, and NIST AI 100-2e2025 \citep{owasp2025llm,mitreatlas2024,vassilev2025adversarial}. We coded representative papers along the following fields:
\begin{itemize}[leftmargin=*]
    \item \textbf{Lifecycle stage:} data, pretraining, alignment, packaging/supply chain, RAG/memory, prompt/inference, tool/agent execution, deployment/maintenance.
    \item \textbf{Security objective:} confidentiality, integrity, availability, safety, privacy, fairness, accountability, agency control.
    \item \textbf{Attacker capability:} black-box prompting, query access, white-box access, data-injection access, tool-description control, RAG-corpus write access, fine-tuning access, supply-chain access, insider access, or deployment access.
    \item \textbf{Attack mechanism:} injection, evasion, jailbreak, extraction, poisoning, backdoor, inversion, membership inference, tool misuse, retrieval manipulation, prompt leakage, or resource exhaustion.
    \item \textbf{Defense layer:} data governance, model training, prompt/context management, retrieval controls, privilege isolation, output verification, monitoring, red teaming, privacy engineering, supply-chain assurance, or incident response.
\end{itemize}

This protocol intentionally avoids treating attack names as mutually exclusive categories. Many real attacks are compositional. For example, an indirect prompt injection can be delivered through a poisoned RAG document, hidden with Unicode obfuscation, cause prompt leakage, and invoke an over-privileged tool. A lifecycle coding makes such composition explicit.

\begin{table*}[t]
\centering
\small
\caption{Attacker models for lifecycle and application-stack analysis.}
\label{tab:attacker-models}
\begin{tabularx}{\textwidth}{L{2.8cm} Y Y Y}
\toprule
\textbf{Attacker model} & \textbf{Access and knowledge} & \textbf{Typical attacks} & \textbf{Relevant defenses} \\
\midrule
Black-box prompt user & Can send prompts and observe outputs; may know public model family but not hidden prompts or weights & jailbreaks, prompt leakage, social-engineering prompts, resource exhaustion & refusal training, rate limits, output filtering, abuse monitoring, prompt-injection detectors \\
\hline
Adaptive query adversary & Can submit many queries and optimize based on responses & adversarial suffix search, model extraction, privacy probing, transfer attacks & query throttling, anomaly detection, randomized defenses, privacy auditing \\
\hline
External-content adversary & Controls content that may later be retrieved or summarized, such as webpages, PDFs, emails, tickets, or repositories & indirect prompt injection, RAG poisoning, malicious tool observations & provenance labels, source allowlists, context isolation, retrieval filtering, least privilege \\
\hline
Corpus/data adversary & Can influence pretraining, instruction-tuning, preference, or evaluation data & data poisoning, backdoors, benchmark contamination, alignment poisoning & data provenance, deduplication, dataset signing, poisoning scans, held-out private evaluations \\
\hline
Fine-tuning/adaptation adversary & Can provide adapters, fine-tuning data, or model updates & safety degradation, hidden triggers, policy drift & safety regression tests, adapter scanning, signed updates, restricted fine-tuning APIs \\
\hline
Supply-chain adversary & Can tamper with model artifacts, tokenizers, packages, plug-ins, or tool manifests & artifact compromise, malicious loaders, compromised plug-ins, tokenizer attacks & signed artifacts, sandboxing, SBOM/MBOM, dependency scanning, reproducible builds \\
\hline
Insider or operator & Has privileged access to logs, prompts, data stores, model versions, or deployment settings & data exfiltration, unsafe policy changes, logging abuse, rollback prevention & separation of duties, audit logs, access control, approval workflows, key management \\
\bottomrule
\end{tabularx}
\end{table*}

\subsection{Terminology}

\textbf{Vulnerability} denotes a weakness in the design, implementation, integration, or operation of an LLM system that can be exploited or accidentally triggered. \textbf{Threat} denotes a potential harmful event enabled by a vulnerability. \textbf{Attack} denotes an intentional action by an adversary. \textbf{Risk} combines likelihood, exposure, and impact. \textbf{Defense} denotes a preventive, detective, corrective, or compensating control.

We distinguish \textbf{jailbreaking} from \textbf{prompt injection}. Jailbreaking usually refers to inducing a model to violate safety policies. Prompt injection refers to causing the model or application to treat attacker-controlled content as instructions. Jailbreaking can be a subclass or goal of prompt injection, but prompt injection also includes data exfiltration, tool misuse, goal hijacking, and prompt leakage \citep{perez2022ignore,greshake2023not,liu2023prompt}. We use \textbf{indirect prompt injection} for attacks where malicious instructions are embedded in external content, such as a document, webpage, email, code repository, tool result, or retrieval item, and later enter the prompt through application logic.

\subsection{Attacker models}

A useful taxonomy must separate the attack surface from the attacker's capability. We use the attacker models in Table~\ref{tab:attacker-models}. These models are not mutually exclusive; a realistic adversary may start as a black-box user, poison a public webpage, wait for it to be retrieved, and then use the resulting agent behavior to obtain additional access.



\section{Lifecycle and Application-Stack Taxonomy}

\begin{figure*}[t]
\centering
\resizebox{0.96\textwidth}{!}{%
\begin{tikzpicture}[
    font=\small,
    stage/.style={rounded corners=2pt, draw=black!55, very thick, minimum height=9mm, align=center, fill=blue!6},
    layer/.style={rounded corners=2pt, draw=black!45, thick, minimum height=7mm, align=center, fill=gray!8},
    risk/.style={rounded corners=2pt, draw=black!45, thick, minimum height=6mm, align=center, fill=orange!10},
    arrow/.style={-Latex, very thick, draw=black!55}
]
\node[layer, minimum height=1.5cm, minimum width=23cm] (stack) {Application stack: user interface $\rightarrow$ prompt/session manager $\rightarrow$ RAG/memory $\rightarrow$ model/runtime $\rightarrow$ tools/agents $\rightarrow$ monitoring/governance};
\node[stage, below=10mm of stack.west, anchor=west, minimum width=2.0cm] (s1) {Data\\collection};
\node[stage, right=5mm of s1, minimum width=2.4cm] (s2) {Pretraining};
\node[stage, right=5mm of s2, minimum width=2.4cm] (s3) {Post-training\\alignment};
\node[stage, right=5mm of s3, minimum width=2.4cm] (s4) {Packaging\\supply chain};
\node[stage, right=5mm of s4, minimum width=2.4cm] (s5) {RAG and\\memory};
\node[stage, right=5mm of s5, minimum width=2.4cm] (s6) {Prompting and\\inference};
\node[stage, right=5mm of s6, minimum width=2.4cm] (s7) {Tool/agent\\execution};
\node[stage, right=5mm of s7, minimum width=2.4cm] (s8) {Deployment and\\maintenance};
\draw[arrow] (s1) -- (s2); \draw[arrow] (s2) -- (s3); \draw[arrow] (s3) -- (s4); \draw[arrow] (s4) -- (s5); \draw[arrow] (s5) -- (s6); \draw[arrow] (s6) -- (s7); \draw[arrow] (s7) -- (s8);
\node[risk, below=8mm of s1, minimum height=0.9cm, minimum width=2.4cm] (r1) {poisoning\\PII leakage};
\node[risk, below=8mm of s2, minimum height=0.9cm, minimum width=2.4cm] (r2) {memorization\\backdoors};
\node[risk, below=8mm of s3, minimum height=0.9cm, minimum width=2.4cm] (r3) {alignment\\poisoning};
\node[risk, below=8mm of s4,minimum height=0.9cm,  minimum width=2.4cm] (r4) {model/package\\tampering};
\node[risk, below=8mm of s5, minimum height=0.9cm, minimum width=2.4cm] (r5) {retrieval\\poisoning};
\node[risk, below=8mm of s6, minimum height=0.9cm, minimum width=2.4cm] (r6) {prompt injection\\jailbreak};
\node[risk, below=8mm of s7, minimum height=0.9cm, minimum width=2.4cm] (r7) {excessive agency\\tool misuse};
\node[risk, below=8mm of s8, minimum height=0.9cm, minimum width=2.4cm] (r8) {monitoring gaps\\availability};
\foreach \a/\b in {s1/r1,s2/r2,s3/r3,s4/r4,s5/r5,s6/r6,s7/r7,s8/r8}{\draw[-Latex, thick, draw=black!35] (\a) -- (\b);}
\node[layer, below=9.6mm of r1.west, anchor=west, minimum height=1.0cm, minimum width=23cm, fill=green!8] (def) {Cross-cutting controls: provenance, least privilege, context isolation, structured outputs, human approval, red teaming, monitoring, incident response};
\end{tikzpicture}%
}
\caption{Lifecycle and application-stack view of LLM vulnerabilities. The same attack technique can appear at multiple stages, but its feasibility, impact, and defensibility depend on where trust boundaries, data flows, and delegated authority are located.}
\label{fig:lifecycle-stack}
\end{figure*}
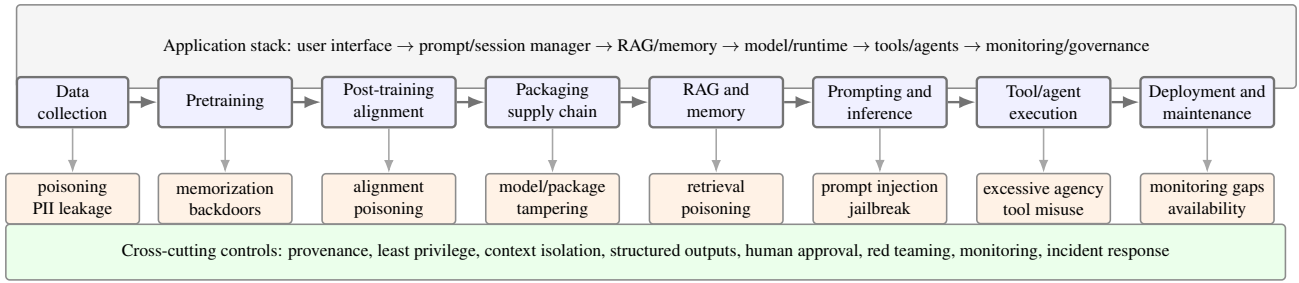


Figure~\ref{fig:lifecycle-stack} illustrates our organizing view. The lifecycle stages are not merely chronological; they are also security boundaries. Data collection determines what the model may memorize or normalize. Pretraining compresses public and private corpora into parameters. Post-training alignment changes refusal behavior and instruction following. Packaging and distribution determine whether users can trust weights, adapters, tokenizer files, and inference code. RAG and memory create mutable knowledge surfaces. Prompting and inference expose the model to direct adversarial interaction. Tool and agent execution delegates authority. Deployment and maintenance determine monitoring, rollback, patching, and incident response.

\begin{table*}[t]
\centering
\small
\caption{Lifecycle-stage view of LLM vulnerabilities, representative risks, and defense families. The table is intended as a systems checklist rather than an exhaustive paper catalogue.}
\label{tab:lifecycle-matrix}
\begin{tabularx}{\textwidth}{L{2.4cm} Y Y Y}
\toprule
\textbf{Stage} & \textbf{Typical vulnerabilities} & \textbf{Representative risks} & \textbf{Defense families} \\
\midrule
Data collection and curation & PII in corpora, biased sources, low-quality or duplicated data, malicious documents, weak provenance & memorization, privacy leakage, data poisoning, representational harms & data documentation, deduplication, PII filtering, provenance tracking, dataset access control, poisoning scans \\
\hline
Pretraining & memorization of rare strings, backdoor triggers, insufficient data isolation, weak privacy accounting & training-data extraction, latent backdoors, unauthorized knowledge retention & privacy-aware training, deduplication, canary tests, backdoor detection, secure training infrastructure \\
\hline
Post-training alignment & poisoned instruction data, unsafe preference data, reward hacking, fine-tuning that removes safeguards & alignment degradation, policy bypass, hidden backdoors, over-refusal or under-refusal & dataset auditing, red-team SFT/RLHF/DPO, alignment regression tests, safety-preserving fine-tuning \\
\hline
Packaging and supply chain & tampered weights, malicious adapters, unsafe model code, tokenizer manipulation, vulnerable dependencies & model compromise, remote code execution, hidden triggers, supply-chain attacks & signed artifacts, reproducible builds, sandboxed loaders, dependency scanning, model cards, software bills of materials \\
\hline
RAG and memory & poisoned retrieval corpora, stale documents, prompt-like content in retrieved data, embedding leakage, cross-user memory contamination & indirect prompt injection, retrieval hijacking, sensitive disclosure, misinformation & source allowlists, provenance labels, retrieval-time filtering, context separation, memory isolation, freshness checks \\
\hline
Prompting and inference & direct injection, jailbreak, adversarial suffixes, prompt leakage, side-channel prompting, output hallucination & safety-policy bypass, developer-prompt disclosure, data exfiltration, harmful content & prompt hardening, input normalization, instruction hierarchy, output verification, refusal calibration, rate limits \\
\hline
Tool and agent execution & excessive privileges, untrusted tool outputs, weak schemas, cross-tool confused deputy, persistent state corruption & unauthorized action, file/database modification, credential leakage, multi-agent propagation & least privilege, capability scoping, structured tool APIs, human approval, sandboxing, transaction logs \\
\hline
Deployment and maintenance & weak monitoring, insecure updates, exposed endpoints, no rollback, resource exhaustion, misconfigured logging & availability loss, privacy breach, delayed detection, incident amplification & observability, anomaly detection, access control, patch management, canary deployment, incident response \\
\bottomrule
\end{tabularx}
\end{table*}

Table~\ref{tab:lifecycle-matrix} shows why a lifecycle view matters. A single term such as ``poisoning'' is insufficient because the object being poisoned changes: pretraining data, alignment data, RAG corpora, tool descriptions, memory, evaluation benchmarks, or monitoring logs. Likewise, a single term such as ``prompt injection'' is insufficient because the payload can be supplied directly by the user, indirectly by retrieved content, persistently through stored memory, or transitively through tool outputs.

\subsection{Security objectives}

Traditional security uses confidentiality, integrity, and availability. LLM systems require these objectives but also extend them.
\begin{itemize}[leftmargin=*]
    \item \textbf{Confidentiality:} prevent unauthorized disclosure of user data, system prompts, proprietary documents, credentials, hidden chain instructions, and training data.
    \item \textbf{Integrity:} preserve correctness of model behavior, retrieved context, tool arguments, outputs, logs, and persistent memories.
    \item \textbf{Availability:} maintain service quality under prompt floods, long-context resource exhaustion, expensive tool loops, or denial-of-wallet attacks.
    \item \textbf{Safety:} prevent harmful instructions, dangerous recommendations, policy bypass, and unsafe execution.
    \item \textbf{Privacy:} reduce memorization, membership inference, inference about sensitive attributes, and cross-user leakage.
    \item \textbf{Fairness:} avoid disproportionate errors, toxic outputs, stereotyping, and exclusionary service behavior.
    \item \textbf{Accountability:} support auditability, provenance, explainability of actions, rollback, and responsibility assignment.
    \item \textbf{Agency control:} ensure the model cannot exceed user intent, tool privileges, or organizational policy when acting in the world.
\end{itemize}

\begin{table*}[t]
\centering
\small
\caption{Mapping LLM attack families to security objectives and common attacker capabilities. A filled cell indicates a frequent primary impact; many attacks have secondary impacts.}
\label{tab:objectives}
\begin{tabularx}{\textwidth}{L{3.7cm} L{3.1cm} c c c c c c c c}
\toprule
\textbf{Attack family} & \textbf{Common capability} & \rotatebox{30}{Conf.} & \rotatebox{30}{Integ.} & \rotatebox{30}{Avail.} & \rotatebox{30}{Safety} & \rotatebox{30}{Privacy} & \rotatebox{30}{Fairness} & \rotatebox{30}{Account.} & \rotatebox{30}{Agency} \\
\midrule
Training-data extraction & query access; memorization probes & $\bullet$ & & & & $\bullet$ & & & \\
Membership inference & query or confidence access & $\bullet$ & & & & $\bullet$ & & & \\
Model extraction & API access, distillation budget & $\bullet$ & $\bullet$ & & & & & & \\
Poisoning/backdoors & data or fine-tuning access & & $\bullet$ & & $\bullet$ & & $\bullet$ & & \\
Jailbreak/adversarial suffix & black-box prompt access & & $\bullet$ & & $\bullet$ & & & & \\
Direct prompt injection & user input control & $\bullet$ & $\bullet$ & & $\bullet$ & $\bullet$ & & & $\bullet$ \\
Indirect prompt injection & external content control & $\bullet$ & $\bullet$ & & $\bullet$ & $\bullet$ & & $\bullet$ & $\bullet$ \\
RAG poisoning & corpus or source control & $\bullet$ & $\bullet$ & & $\bullet$ & $\bullet$ & & $\bullet$ & \\
Tool misuse/excessive agency & prompt or tool-output control & $\bullet$ & $\bullet$ & $\bullet$ & $\bullet$ & $\bullet$ & & $\bullet$ & $\bullet$ \\
Resource exhaustion & query and context budget & & & $\bullet$ & & & & & \\
Supply-chain compromise & artifact/dependency control & $\bullet$ & $\bullet$ & $\bullet$ & $\bullet$ & $\bullet$ & & $\bullet$ & $\bullet$ \\
\bottomrule
\end{tabularx}
\end{table*}

\section{Attack-Family Synthesis}

The lifecycle taxonomy answers where a vulnerability appears. This section synthesizes the major attack families across stages. The goal is to connect mechanisms that are often studied separately. Table \ref{tab:objectives} maps the LLM attack families to security objectives and common attacker capabilities, and Table \ref{tab:attack-synthesis} summarizes attack-family synthesis across lifecycle stages.

\subsection{Privacy leakage, extraction, and inference}

Privacy attacks exploit the fact that LLMs are trained on, adapted with, or connected to sensitive information. Training-data extraction recovers memorized text from model completions, especially when sequences are rare, duplicated, or overrepresented \citep{carlini2021extracting}. Membership inference estimates whether a candidate sample participated in training \citep{shokri2017membership,nasr2019comprehensive}. Model inversion attempts to reconstruct representative inputs or sensitive attributes \citep{fredrikson2015model}. In RAG systems, privacy leakage may occur even if the base model is safe: a user may retrieve documents they should not access, an embedding may reveal information through nearest-neighbor behavior, or a prompt injection may cause an agent to summarize private context.

Privacy evaluation must therefore specify the data location. Leakage from weights, context windows, retrieval stores, logs, telemetry, memories, and tool outputs require different mitigations. A memorization defense will not protect against over-permissive retrieval. A retrieval access-control policy will not protect against secrets placed directly inside system prompts. This is why privacy should be evaluated as an end-to-end property of the application stack.

\subsection{Adversarial prompting, jailbreaks, and evasion}

Adversarial prompting includes direct prompt injection, jailbreaks, adversarial suffixes, role-play attacks, multilingual attacks, encoding/obfuscation, and multi-turn persuasion \citep{perez2022ignore,wallace2019universal,zou2023universal,wei2023jailbroken}. These attacks exploit the model's instruction-following objective and the ambiguity of natural language. The most important distinction is between attacks that merely elicit unsafe text and attacks that alter application control flow. In a tool-free chatbot, a jailbreak may produce policy-violating content. In a connected agent, the same jailbreak can become a command to read, write, send, or execute.

Evasion also targets safety classifiers and guard models. Attackers may change formatting, scripts, Unicode characters, word boundaries, or language to bypass filters while preserving meaning. Such attacks demonstrate that lexical filtering alone is fragile. Robust systems combine normalization, semantic detection, model-level robustness, and post-generation action controls.

\subsection{Prompt leakage and system-prompt exposure}

Prompt leakage attempts to reveal hidden instructions, policies, examples, tool descriptions, or developer prompts. The security impact depends on what the prompt contains. If the prompt contains only general behavioral guidance, leakage may be low impact. If it contains credentials, private data, proprietary workflows, or security policies that enable bypass, leakage becomes a confidentiality and integrity risk. Good design minimizes secrets in prompts and treats prompts as potentially recoverable configuration, not secure storage.

\subsection{Poisoning, backdoors, and alignment attacks}

Poisoning alters data or feedback so that the model learns attacker-preferred behavior. Backdoors are a special case in which the model behaves normally except when triggered \citep{gu2017badnets,chen2017targeted,kurita2020weight}. In LLMs, poisoning can occur at pretraining, instruction tuning, preference optimization, RAG ingestion, memory updates, tool descriptions, or benchmark construction. Alignment attacks target the post-training stage by poisoning safety examples or weakening refusal behavior during fine-tuning \citep{wan2023poisoning,qi2023fine}. The clean-performance constraint makes these attacks difficult to detect: the model may pass standard benchmarks while failing on trigger contexts.

Defenses require provenance and regression testing. Data should be traceable; safety tests should include trigger-like contexts; adapters should be scanned before merging; and fine-tuned models should be compared against their base models on safety, privacy, and utility. Benchmark contamination must also be monitored because LLMs may memorize public evaluation sets, leading to overestimated robustness.

\subsection{RAG, vector databases, and memory attacks}

RAG attacks manipulate what the model sees at inference time. Retrieval poisoning inserts documents that are likely to be retrieved for target queries. Indirect prompt injection embeds instructions inside external content. Vector-store attacks may exploit embedding similarity, chunk boundaries, stale indexes, or missing access control \citep{xiong2024ragsecurity,zhao2024ragpoisoning}. Memory attacks persist false or malicious state across sessions.

The key insight is that RAG turns content security into instruction security. A document is no longer merely read by a human; it is parsed by a model that may treat it as a command. The defense is not to ban RAG, but to preserve source trust, label untrusted content, and prevent retrieved text from authorizing actions.

\subsection{Tool, plug-in, and agent attacks}

Agents amplify LLM failures because they connect model outputs to external effects. Tool attacks include malicious tool outputs, prompt injection through API responses, unsafe command construction, tool-description poisoning, excessive permissions, and cross-tool data exfiltration \citep{ferrag2025agentthreats,ling2026secureagents,zhao2026clawguard}. Multi-agent systems add propagation risks: one compromised agent can influence another through messages, shared memory, or delegated tasks.

The most dangerous pattern is the confused deputy. The user delegates authority to the agent; the agent reads attacker-controlled content; the content persuades the agent to use the user's authority for the attacker's goal. Preventing this requires explicit authority tracking. The system should know which principal authorized each action and which untrusted content influenced it.

\subsection{Availability, cost, and abuse}

LLM availability attacks include prompt floods, long-context resource exhaustion, retrieval amplification, expensive tool loops, recursive agent calls, and denial-of-wallet attacks. Abuse also includes using LLMs to scale phishing, spam, misinformation, malware assistance, or vulnerability discovery. While these harms are not always vulnerabilities in the model itself, they are risks of deployment. Defenses include quotas, abuse detection, proof-of-work or payment controls for high-volume access, tool-call limits, circuit breakers, and misuse monitoring.

\begin{table*}[t]
\centering
\small
\caption{Attack-family synthesis across lifecycle stages. The same family can appear at multiple layers with different defenses.}
\label{tab:attack-synthesis}
\begin{tabularx}{\textwidth}{L{2.8cm} Y Y Y}
\toprule
\textbf{Family} & \textbf{Lifecycle manifestations} & \textbf{Primary failure mode} & \textbf{Most useful controls} \\
\midrule
Privacy and extraction & pretraining memorization, prompt leakage, RAG over-retrieval, log exposure, memory leakage & confidential information becomes accessible through generation or retrieval & privacy-aware training, access control, secret minimization, log redaction, memory isolation \\
Prompt injection and jailbreak & direct user prompts, indirect retrieved content, tool outputs, multimodal inputs, stored memories & untrusted language is interpreted as instruction & context isolation, instruction hierarchy, detectors, least privilege, output/action verification \\
Poisoning and backdoors & data collection, instruction tuning, preference data, RAG corpora, adapters, tool manifests & attacker changes future behavior while preserving normal utility & provenance, data signing, anomaly scans, trigger tests, safety regression suites \\
Supply-chain compromise & model files, adapters, tokenizers, inference code, dependencies, plug-ins & trusted artifact contains malicious behavior or vulnerable code & signed artifacts, sandboxed loading, dependency scanning, MBOM/SBOM, reproducible builds \\
Agent/tool misuse & tool schemas, API calls, code execution, email/browser/file access, multi-agent messages & model output causes unauthorized external effect & capability scoping, human approval, dry-run mode, transaction logs, policy engines \\
Availability and cost abuse & long context, repeated queries, retrieval/tool loops, expensive generation, API amplification & attacker consumes resources or degrades service & rate limits, quotas, timeouts, circuit breakers, billing anomaly detection \\
\bottomrule
\end{tabularx}
\end{table*}

\section{Threats Across the Lifecycle}

\subsection{Data collection and curation}

Large-scale pretraining relies on web crawls, code repositories, books, forums, scientific articles, product documentation, synthetic data, and human-generated instruction examples. The main vulnerabilities are weak provenance, unknown consent, duplication, private information, toxic or biased content, and malicious data insertion. The data stage creates both model-level and system-level risks. A model may memorize rare strings or private records \citep{carlini2019secret,carlini2021extracting}; it may reproduce stereotypes or harmful associations \citep{bender2021dangers,weidinger2021ethical}; or it may learn trigger-response behavior if attackers insert poisoned documents before training \citep{gu2017badnets,chen2017targeted,kurita2020weight}.

A lifecycle perspective distinguishes passive contamination from active poisoning. Passive contamination includes accidental inclusion of personal information, outdated facts, or biased sources. Active poisoning includes attacker-selected text that induces a future behavior, such as a backdoor trigger, malicious code pattern, or false association. LLM data scale does not eliminate this risk; scale can dilute triggers, but it also makes exhaustive manual auditing infeasible. Synthetic-data pipelines add another complication: if model-generated content re-enters training data without provenance, errors and attack artifacts can become self-reinforcing.

Defenses at this stage should be preventive. They include dataset documentation, source reputation scoring, near-duplicate removal, PII detection, toxicity screening, canary insertion for memorization tests, poisoning anomaly detection, and access control on data pipelines. These defenses have tradeoffs: aggressive filtering can remove minority dialects or specialized technical content, while weak filtering can preserve private or malicious data. The research gap is not only better filters, but auditable data governance that can answer which sources influenced which model behavior.

\subsection{Pretraining and model internals}

Pretraining compresses large corpora into parameters. This compression can create privacy leakage and hidden integrity failures. Extraction attacks exploit model completions to recover memorized training sequences, especially rare or duplicated strings \citep{carlini2021extracting}. Membership inference attacks attempt to determine whether a sample was present in training \citep{shokri2017membership,nasr2019comprehensive}. Model inversion aims to infer properties or representative inputs from outputs or internal representations \citep{fredrikson2015model}. Model extraction and distillation attacks aim to replicate functionality through API queries \citep{tramer2016stealing}.

Backdoors and weight poisoning are integrity threats. In a backdoored model, behavior appears normal except when a trigger activates malicious behavior \citep{gu2017badnets,kurita2020weight}. For LLMs, triggers can be lexical, syntactic, semantic, multilingual, or embedded in code comments. Unlike image classifiers, LLMs may execute backdoor behavior through long-form generation or tool calls, so evaluation must measure not only class accuracy but downstream action.

Mitigations include deduplication, privacy-aware training, differential privacy, controlled memorization tests, gradient clipping, red-team prompts, backdoor scanning, and model-card disclosure \citep{dwork2006differential,abadi2016deep,carlini2019secret}. Differential privacy can reduce memorization but may degrade utility at large model scale if not carefully engineered. Backdoor detection remains difficult when triggers are semantic rather than token-level. A practical defense should combine data curation, training-time tests, inference-time monitoring, and limited authority for downstream actions.

\subsection{Post-training alignment and fine-tuning}

Instruction tuning, RLHF, RLAIF, constitutional AI, and DPO improve helpfulness and safety but also introduce new attack surfaces \citep{ouyang2022training,ziegler2019fine,stiennon2020learning,bai2022constitutional,rafailov2023direct}. Alignment datasets can be poisoned, preference models can encode hidden policies, and downstream fine-tuning can weaken refusal behavior \citep{wan2023poisoning,qi2023fine}. The alignment stage is therefore both a defense and a target.

Post-training vulnerabilities have three common forms. First, poisoned examples can teach a model to comply with unsafe requests under specific contexts. Second, preference optimization can overfit to superficial refusal templates, making the model brittle under rephrasing, role play, or multi-turn pressure. Third, benign fine-tuning for domain adaptation can erode safety boundaries because the optimization objective rewards task compliance rather than safety preservation.

Defenses should treat safety as a regression property. Any fine-tuning or adapter update should be tested against a stable suite of safety, privacy, jailbreak, and task-utility prompts. Safety-preserving fine-tuning should freeze or regularize parts of the model when possible, include refusal and boundary examples, and evaluate cross-domain transfer. The field still lacks reliable certificates that an aligned model remains aligned after adapter merging, quantization, distillation, or local fine-tuning.

\subsection{Packaging, distribution, and supply chain}

Open models, adapters, tokenizers, inference servers, model-loading scripts, and third-party packages create a software supply chain. LLM systems frequently load artifacts from public model hubs, run custom Python code, and combine multiple dependencies. Vulnerabilities include malicious model files, unsafe deserialization, compromised adapters, tokenizer manipulation, dependency confusion, prompt templates hidden in packages, and contaminated evaluation scripts.

The risk is amplified because model artifacts are large, opaque, and hard to inspect. A malicious adapter can change behavior without modifying base weights. A tokenizer can alter how dangerous strings are segmented. A model repository can include code that executes at load time. Supply-chain attacks can also occur through plug-ins, tool manifests, browser extensions, MCP-like servers, and agent skills \citep{ferrag2025agentthreats,chu2026systematicagents,dehghantanha2026sok}.

Defenses should import mature software-security practices into ML operations: signed artifacts, hashes, reproducible builds, sandboxed model loading, dependency scanning, software bills of materials, model bills of materials, vulnerability disclosure, least-privilege inference containers, and provenance labels. Model cards should be extended to security cards that disclose training data classes, alignment methods, tool privileges, known limitations, and intended deployment boundaries.

\subsection{Retrieval-augmented generation and memory}

RAG connects an LLM to external knowledge through retrieval and context construction \citep{lewis2020retrieval,karpukhin2020dense}. It reduces some hallucinations and supports private-domain applications, but it introduces a new pathway by which untrusted data becomes model context. A retrieved document can contain instructions that override the user's goal, leak system prompts, exfiltrate information, or steer summaries. This is the canonical setting for indirect prompt injection \citep{greshake2023not,liu2023prompt}. Retrieval can also be poisoned by adding documents that dominate nearest-neighbor search, manipulate embeddings, or appear authoritative \citep{xiong2024ragsecurity,zhao2024ragpoisoning}.

RAG creates three distinct trust problems. First, retrieved content has uncertain provenance. Second, retrieved content is mixed with developer instructions inside the model context. Third, retrieved content is often treated as evidence even when it is stale, adversarial, or out of distribution. Persistent memory adds a temporal dimension: a malicious or mistaken memory can affect future sessions, and cross-user memory bugs can become privacy breaches.

Defenses include source allowlists, retrieval-time safety classification, provenance metadata, freshness checks, user-visible citations, context partitioning, and memory isolation. The model should not be asked to infer which text is instruction and which text is data solely from natural-language formatting. Application logic should label untrusted content, limit what retrieved content can authorize, and verify claims against multiple sources when high impact. Research is needed on retrieval systems that preserve provenance through embeddings and context construction rather than losing it at the vector-search boundary.

\subsection{Prompting and inference}

Prompting is the most visible attack surface. Direct prompt injection attempts to hijack goals, reveal hidden prompts, bypass policies, or override previous instructions \citep{perez2022ignore}. Jailbreaks and adversarial suffixes exploit instruction-following behavior and transfer across models \citep{zou2023universal,wei2023jailbroken}. Universal adversarial triggers show that small textual patterns can systematically change model behavior \citep{wallace2019universal}. Programmatic behavior can also be exploited when models process code-like or instruction-like structures \citep{kang2023exploiting}.

The root cause is not simply that models are insufficiently aligned. The root cause is that natural language is used simultaneously as data, instruction, policy, evidence, and executable plan. Delimiters, system prompts, and refusal templates help but do not create cryptographic separation. Models can be confused by role-play, translation, encoding, multi-turn pressure, conflicting instructions, hidden Unicode, or tool outputs. Larger context windows increase the volume of untrusted content that may affect behavior.

Defenses include input normalization, Unicode and encoding governance, prompt templates that separate roles, instruction hierarchy, prompt-injection detectors, output safety filters, rate limiting, and robust refusal training. However, these defenses are probabilistic when they rely on the model to classify intent. Deterministic controls should therefore constrain what the model can do after generation. For example, the model may propose a database query, but a schema validator and policy engine should decide whether the query can run.

\subsection{Tools, plug-ins, and agents}

Tool-using LLMs convert language into actions. ReAct-style prompting, Toolformer-like tool use, API-augmented models, and agent frameworks allow models to plan, retrieve, compute, call APIs, and update state \citep{yao2022react,schick2023toolformer,qin2023toolllm}. This creates new risks: excessive agency, tool-output prompt injection, confused-deputy behavior, insecure tool schemas, credential leakage, multi-step escalation, and persistent state corruption. In security operations, LLMs may help triage vulnerabilities or automate penetration-testing tasks, but this also increases dual-use concerns \citep{deng2023pentestgpt}.

Agent security differs from chatbot security in three ways. First, agents possess delegated authority. A harmful output may become an email, transaction, code commit, database update, or shell command. Second, agents are stateful. They store memories, partial plans, and tool observations across steps. Third, agents interact with untrusted environments. Web pages, repositories, tickets, emails, and PDFs can contain instructions targeting the agent rather than the human user. Recent work therefore treats LLM-agent ecosystems as protocol and workflow security problems, not only model-safety problems \citep{ferrag2025agentthreats,ling2026secureagents,chu2026systematicagents,zhao2026clawguard,dehghantanha2026sok}.

Defenses must control authority explicitly. Tools should be least-privilege, scoped to the user's current task, and separated by sensitivity. Tool calls should use typed schemas, not free-form strings. High-impact operations should require human approval. Tool outputs should be treated as untrusted data and routed through the same context-isolation and normalization pipeline as web content. Sandboxing, dry-run modes, transaction logs, and reversible execution reduce impact. A central research challenge is compositionality: a system can have safe components but unsafe interactions when a model chains them.

\subsection{Deployment, monitoring, and maintenance}

Deployment determines whether failures are detected, contained, and corrected. LLM applications can fail through exposed endpoints, weak authentication, logs containing secrets, excessive context costs, denial-of-wallet attacks, prompt floods, tool loops, unbounded recursion, model-version drift, stale safety prompts, and missing incident response. Maintenance adds update channels, rollback procedures, dependency patching, prompt-template changes, and red-team regression tests.

Availability is particularly under-discussed. Long-context models can be expensive; attackers may send inputs that maximize token usage, retrieval load, or tool calls. Agents can enter loops or repeatedly call costly APIs. Defensive design should include token budgets, tool-call quotas, timeouts, circuit breakers, queue isolation, and billing anomaly detection. Monitoring should not only log prompts and completions; it should log provenance, retrieved document IDs, tool arguments, policy decisions, user approvals, model version, and safety-detector outputs.

Incident response for LLM systems differs from conventional software response. A fix may involve changing prompts, safety policies, retrieval filters, memory entries, model versions, adapters, tool permissions, or data sources. Some incidents require forgetting or quarantining memories; others require retraining or revoking model artifacts. Organizations therefore need LLM-specific runbooks for prompt-injection incidents, data leakage, unsafe tool use, poisoning discovery, and model rollback.

\section{Defense-in-Depth Architecture}

\begin{figure*}[t]
\centering
\resizebox{0.92\textwidth}{!}{%
\begin{tikzpicture}[
  font=\small,
  box/.style={rounded corners=2pt, draw=black!55, thick, minimum width=2.9cm, minimum height=8mm, align=center, fill=blue!5},
  guard/.style={rounded corners=2pt, draw=black!55, thick, minimum width=3.1cm, minimum height=8mm, align=center, fill=green!10},
  red/.style={rounded corners=2pt, draw=black!55, thick, minimum width=3.1cm, minimum height=8mm, align=center, fill=orange!10},
  arrow/.style={-Latex, very thick, draw=black!60}
]
\node[box] (u) {User and\\external content};
\node[guard, right=5mm of u] (norm) {Input hygiene\\and provenance};
\node[guard, right=5mm of norm] (ctx) {Context isolation\\and policy routing};
\node[box, right=5mm of ctx] (llm) {LLM / planner};
\node[guard, right=5mm of llm] (out) {Output schema\\and action verifier};
\node[red, right=5mm of out] (tool) {Tools, APIs,\\and data stores};
\node[guard, below=8mm of ctx] (mon) {Telemetry, anomaly detection, \\ red-team regression tests};
\node[guard, below=8mm of out] (hitl) {Human approval for irreversible or \\ high-impact actions};
\draw[arrow] (u) -- (norm); \draw[arrow] (norm) -- (ctx); \draw[arrow] (ctx) -- (llm); \draw[arrow] (llm) -- (out); \draw[arrow] (out) -- (tool);
\draw[arrow] (ctx) -- (mon); \draw[arrow] (out) -- (hitl); \draw[arrow] (hitl) -- (tool);
\draw[-Latex, thick, dashed, draw=black!45] (tool.north) to[out=150,in=15] node[above, align=center] {tool observations are untrusted inputs} (norm.north);
\end{tikzpicture}%
}
\caption{Defense-in-depth architecture for LLM applications. Reliable mitigation cannot rely only on model behavior; deterministic controls should constrain information flow and delegated authority before and after the model call.}
\label{fig:defense-architecture}
\end{figure*}
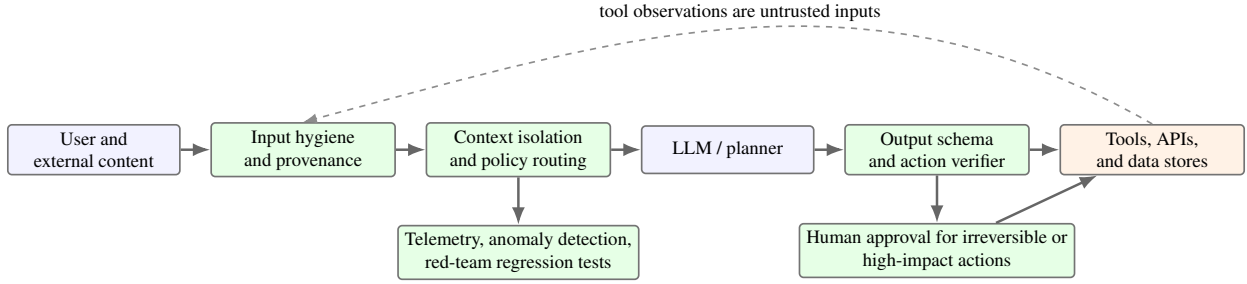


No single defense is sufficient. Model alignment reduces risk but cannot guarantee that the model will correctly separate data from instructions in all contexts. Prompt filtering catches some attacks but can be bypassed or over-block benign use. RAG citations improve transparency but do not prevent malicious retrieved content from influencing behavior. Tool schemas reduce injection into APIs but not necessarily goal hijacking. Therefore, the core design principle is defense in depth: combine deterministic controls that bound actions with probabilistic controls that detect and reduce residual risk. Figure \ref{fig:defense-architecture} summarizes a general defense-in-depth architecture for LLM applications. 

\subsection{Deterministic controls}

Deterministic controls should be preferred wherever a security property can be expressed outside the model. Examples include least-privilege tool permissions, allowlisted network destinations, typed API schemas, JSON schema validation, sandboxed code execution, read-only defaults, transaction limits, human approval for irreversible operations, retrieval-source allowlists, context-size budgets, and cryptographic signing of model artifacts. These controls are auditable and do not require the model to reason perfectly under adversarial input.

A useful rule is that untrusted natural language should not directly authorize external actions. It may provide evidence, a candidate plan, or a draft, but application logic should verify whether the requested action is within scope. For example, a retrieved document may say that an agent should send a file to a URL, but the system should treat that instruction as data and reject it unless the authenticated user independently authorized the action.

\subsection{Model-level and prompt-level controls}

Model-level defenses include adversarial training, refusal calibration, safety fine-tuning, preference optimization, constitutional rules, and robustness evaluation \citep{ganguli2022red,ouyang2022training,bai2022constitutional,rafailov2023direct}. Prompt-level defenses include system prompts, instruction hierarchy, delimiters, reminders about untrusted content, self-check prompts, and chain-of-verification. These controls improve behavior but remain statistical. They should be evaluated continuously against adaptive attacks and regressions.

Prompt hardening is most useful when paired with architectural separation. A prompt may tell the model that retrieved content is untrusted; a system design should also ensure that retrieved content cannot directly set tool permissions, change policy, or erase logs. Similarly, a model may be trained to refuse prompt-leakage requests; the application should still avoid placing unnecessary secrets in prompts.

\subsection{Retrieval and memory controls}

RAG defenses should operate before, during, and after retrieval. Before retrieval, systems should curate sources, scan documents, attach provenance, and limit ingestion rights. During retrieval, systems should filter malicious-looking content, diversify sources, and preserve metadata. After retrieval, systems should label context, quote evidence separately from instructions, and verify high-impact claims. Persistent memory should be scoped by user, organization, task, and time; sensitive memory should expire or require explicit confirmation.

Embedding systems need additional controls. Embeddings can leak information, and nearest-neighbor search can be manipulated by adversarial documents. Access control should be enforced before retrieval rather than after the vector search whenever possible. If a user lacks access to a document, its embedding should not influence retrieval results for that user. Provenance should not be discarded when text is chunked and embedded.

\subsection{Tool and agent controls}

Tools should be designed as capability objects rather than broad ambient authority. Each tool should declare its scope, inputs, outputs, side effects, and risk level. Agents should receive only the tools needed for the current task. Sensitive tools should support dry-run, preview, confirmation, and rollback. Tool outputs should be untrusted observations, not instructions. Cross-tool data flow should be mediated by policy: a web-reading tool should not be able to instruct an email-sending tool to exfiltrate private data.

Agent evaluation should include long-horizon attacks, persistent memory attacks, malicious tool outputs, poisoned webpages, compromised plug-ins, and multi-agent propagation. Short single-turn jailbreak benchmarks underestimate risk because many agent failures emerge only after multiple tool calls and state updates.

\subsection{Monitoring, red teaming, and governance}

Monitoring closes the loop. Useful telemetry includes prompt hashes, retrieved source IDs, model version, tool-call arguments, policy decisions, refusal reasons, detector scores, human approvals, and anomalies in token usage or tool-call frequency. Privacy-sensitive logging should use minimization, access control, retention limits, and secure redaction.

Red teaming should be continuous rather than one-time. Regression suites should include known jailbreaks, prompt-injection patterns, RAG poisoning cases, privacy probes, tool misuse scenarios, and availability stress tests \citep{ganguli2022red,zhan2024garak}. Governance should define risk ownership, review processes for new tools, incident-response procedures, and user-facing disclosure. Security evaluations should be versioned because model and prompt updates can change behavior even when application code remains fixed.

\section{Evaluation Methodology}

Evaluation is a central weakness of current LLM-security research. Many papers report attack success rate on a small set of prompts, but deployment risk depends on attacker capability, tool privileges, task distribution, monitoring, and user behavior. A systematization should separate model-level metrics from system-level metrics.

\subsection{Metrics}

Common model-level metrics include attack success rate, refusal rate, harmful completion rate, toxicity score, privacy exposure, exact training-string recovery, membership-inference advantage, backdoor trigger success, clean-task accuracy, and transferability across models. System-level metrics include unsafe action rate, unauthorized data access, privilege-escalation distance, time to detection, rollback success, false positive rate of filters, user-task utility, cost amplification, and incident severity.

Attack success rate should be interpreted carefully. A jailbreak that produces unsafe text is not equivalent to an agent that sends an unauthorized email. Conversely, a low text-level harmfulness score may hide integrity failure if the model silently changes a database query. Security evaluations should therefore specify the protected asset, attacker goal, allowed actions, and success criteria.

\subsection{Benchmarks and test design}

Benchmarks should cover direct prompts, indirect content, RAG corpora, tool outputs, code repositories, multimodal inputs, persistent memory, and multi-turn sessions. They should include benign hard cases to measure over-refusal and utility loss. For RAG and agents, benchmark tasks should include realistic documents, permissions, and tool side effects. AgentDojo-like environments illustrate the importance of evaluating both task success and attack resistance \citep{zhang2024agentdojo,zhao2026clawguard}.

Reproducibility is difficult because many leading models are closed, versioned silently, and protected by changing safety layers. Papers should report model name, date, version, decoding parameters, system prompts when publishable, prompt templates, tool schemas, detector thresholds, and source code. For closed models, repeated evaluation over time is necessary because behavior can drift.

\subsection{Threat-model reporting}

Every evaluation should state attacker knowledge and access. A black-box user who can only submit prompts is different from an attacker who can upload documents to a RAG corpus, control a webpage, edit a tool manifest, fine-tune an adapter, or tamper with model artifacts. The paper should also report whether the attack requires many queries, hidden text, model gradients, training access, or social engineering. Without this reporting, comparisons across defenses are misleading. Table \ref{tab:reporting-checklist} lists our recommended reporting items for LLM vulnerability studies. 

\begin{table*}[h!]
\centering
\small
\caption{Recommended reporting checklist for LLM vulnerability studies.}
\label{tab:reporting-checklist}
\begin{tabularx}{\textwidth}{L{3.3cm} Y Y}
\toprule
\textbf{Item} & \textbf{What to report} & \textbf{Why it matters} \\
\midrule
System boundary & base model, prompt layer, RAG, tools, memory, deployment stack & identifies what is actually being evaluated \\
Attacker capability & prompt access, corpus write access, tool-output control, fine-tuning access, artifact control & determines feasibility and relevance \\
Protected asset & private data, model weights, safety policy, tool action, availability, user trust & avoids vague claims of ``security'' \\
Success criterion & harmful text, data leakage, unauthorized action, cost increase, policy violation & makes attack success comparable \\
Utility metric & task accuracy, helpfulness, latency, cost, over-refusal, user satisfaction & detects defenses that simply break the system \\
Model/version details & model family, date, parameters when known, decoding, safety layer & supports reproduction and drift analysis \\
Context construction & system prompt, retrieved documents, memory policy, delimiters, tool schemas & critical for prompt-injection and RAG studies \\
Defense placement & pre-input, model, retrieval, output, tool boundary, monitoring, human approval & distinguishes probabilistic from deterministic controls \\
Reproducibility & code, prompts, datasets, random seeds, refusal classification rules & enables independent verification \\
Ethical controls & safe examples, disclosure process, harm minimization & reduces dual-use risk \\
\bottomrule
\end{tabularx}
\end{table*}

\section{Risk Prioritization for Deployment}
\label{sec:risk-prioritization}

A key lesson from the lifecycle view is that vulnerabilities should not be prioritized only by their standalone attack success rate. A jailbreak that succeeds against an isolated chatbot may be less operationally dangerous than a lower-success-rate indirect prompt-injection attack against an enterprise agent with access to private documents, email, calendars, code repositories, or production systems. Deployment risk is therefore a function of the \emph{model}, the \emph{application stack}, the \emph{available tools}, the \emph{data context}, and the \emph{authority delegated to the system}. This section provides a practical prioritization framework for translating the vulnerability taxonomy into deployment decisions.

\subsection{A Deployment-Oriented Risk Model}

For deployed \llm{} systems, risk should be evaluated at the level of a \emph{model--application--authority tuple}, rather than at the level of the base model alone. We define a deployment scenario by
\[
\mathcal{S} = (\mathcal{M}, \mathcal{C}, \mathcal{D}, \mathcal{T}, \mathcal{A}, \mathcal{H}),
\]
where $\mathcal{M}$ is the model or model ensemble, $\mathcal{C}$ is the prompt and context-construction pipeline, $\mathcal{D}$ is the set of accessible data sources, $\mathcal{T}$ is the set of tools and external interfaces, $\mathcal{A}$ is the authority delegated to the system, and $\mathcal{H}$ represents human oversight and incident-response mechanisms. This formulation makes explicit why two deployments using the same model can have very different security profiles.

A practical prioritization score can be described as
\[
R = f(E, F, I, D, O),
\]
where $E$ denotes exposure, $F$ attacker feasibility, $I$ impact, $D$ defense maturity, and $O$ observability. Exposure measures how many users, documents, tools, and external channels can reach the system. Feasibility captures the attacker's required knowledge, access, cost, and repeatability. Impact measures confidentiality, integrity, availability, safety, privacy, legal, and reputational consequences. Defense maturity captures whether controls are deterministic, tested, auditable, and resilient to adaptive attackers. Observability measures whether the organization can reconstruct what happened after a security event. This model is consistent with the broader adversarial-machine-learning view that risk depends not only on attack method, but also on attacker capability, system objective, and operational context \citep{vassilev2025adversarial,owasp2025llm,mitreatlas2024}.

\subsection{Risk Drivers Across the LLM Application Stack}

In operational settings, five risk drivers are especially important.

\textbf{External authority.} The most important question is what the system can do outside the chat window. A system that only generates text has limited direct side effects. A system that can send email, modify code, query databases, invoke cloud APIs, make purchases, operate robots, or run terminal commands has substantially higher agency risk. For this reason, tool access and delegated authority are often stronger predictors of risk than model size.

\textbf{Untrusted context.} Indirect prompt injection becomes possible when untrusted content is placed into the model context. This includes retrieved web pages, PDFs, emails, code comments, repository files, calendar descriptions, database records, browser content, tool outputs, and vector-database chunks \citep{greshake2023not,liu2023prompt,xiong2024ragsecurity,zhao2024ragpoisoning}. A deployment that mixes untrusted content with instructions and private data should be treated as high risk unless it uses strong provenance, isolation, and authorization controls.

\textbf{Private and regulated data.} Enterprise assistants, healthcare systems, financial systems, educational platforms, and legal tools often expose models to sensitive data. Risks include training-data memorization, retrieval leakage, embedding leakage, membership inference, prompt logging, cross-tenant data exposure, and unauthorized summarization \citep{carlini2021extracting,shokri2017membership}. The priority of privacy controls increases when data is personally identifiable, legally protected, proprietary, or security-sensitive.

\textbf{Irreversible or high-impact actions.} Some actions are difficult to undo, such as sending confidential information, deleting files, merging code, changing access policies, executing financial transactions, publishing content, or triggering physical movement. These actions require stronger approval, simulation, transaction preview, logging, and rollback mechanisms.

\textbf{Evidence after an incident.} A system that cannot explain which model version, prompt, retrieved document, tool call, policy decision, or human approval produced an action is difficult to audit. Observability is therefore a security control, not only an engineering convenience. Logs should preserve security-relevant events while minimizing unnecessary retention of sensitive content.

\subsection{Scenario-Oriented Risk Prioritization}

Table~\ref{tab:risk-prioritization} summarizes common deployment scenarios. The priority level is not absolute: it should be adjusted according to the sensitivity of the domain, the maturity of controls, and the degree of external authority. However, the table illustrates a recurring pattern: \llm{} risks become most urgent when untrusted content, private data, and external actions are combined.

\begin{table*}[t] \centering \small \caption{Deployment-oriented risk prioritization. Priority depends on the specific application, but this matrix summarizes common patterns.} \label{tab:risk-prioritization} \begin{tabularx}{\textwidth}{L{3.0cm} L{2.4cm} L{2.4cm} Y Y} \toprule \textbf{Scenario} & \textbf{Typical priority} & \textbf{Dominant objective} & \textbf{Why} & \textbf{First controls to implement} \\ \midrule Public chatbot without tools & medium & safety, abuse & large exposure, but limited external side effects & abuse monitoring, refusal calibration, rate limits, safety filters \\ Enterprise RAG assistant over private documents & high & confidentiality, integrity & private data and indirect prompt injection through documents & access control before retrieval, provenance labels, context isolation, logging \\ Coding assistant with repository and terminal access & high & integrity, agency & prompt injection in files can lead to code changes or command execution & sandboxing, least privilege, command approval, repository trust policy \\ Email/calendar/browser agent & high & privacy, agency & untrusted messages or webpages can trigger actions under user authority & tool scoping, human confirmation, untrusted-content handling, transaction logs \\ Fine-tuned domain model & medium--high & safety, privacy & private fine-tuning data and safety regression risk & data audit, privacy tests, safety regression, adapter/version control \\ Open model distribution & high & supply-chain integrity & many downstream users trust artifacts and loaders & signed weights, safe serialization, reproducible release, dependency scanning \\ Security-operation or penetration-testing assistant & high & safety, accountability & dual-use tasks and privileged context & role-based access, audit logs, policy gating, controlled tool environments \\ \bottomrule \end{tabularx} \end{table*}

\subsection{Control Selection by Risk Driver}

Risk prioritization should lead to concrete control selection. The most effective controls are often not better prompts, but stronger system boundaries. Prompt-level instructions are useful but should not be treated as the primary security mechanism when the system has access to private data or external tools.

\subsection{Domain-Specific Considerations}

\textbf{Healthcare.} Healthcare deployments combine high privacy requirements with high consequences for hallucinated or incomplete advice. The most important risks are leakage of protected health information, unsupported medical recommendations, biased triage, unsafe summarization of clinical records, and poor auditability. \rag{} sources should be curated and versioned; outputs should include provenance; and high-impact recommendations should remain under professional oversight. The model should assist documentation, retrieval, and patient communication rather than independently making diagnosis or treatment decisions.

\textbf{Finance and insurance.} Financial systems face risks involving private records, regulatory compliance, fraud, market manipulation, and unauthorized transactions. LLMs used for customer support, claims processing, investment research, or internal analytics should separate advice generation from transaction authority. Controls should include source provenance, suitability checks, human approval for financial actions, data-retention limits, and strong audit trails.

\textbf{Education.} Educational deployments involve student privacy, academic integrity, biased feedback, misinformation, and inappropriate delegation of grading authority. Systems should disclose limitations, avoid storing unnecessary student data, distinguish tutoring from assessment, and preserve instructor oversight. For grading or feedback systems, rubrics, provenance, and appeal mechanisms are important accountability controls.

\textbf{Cybersecurity.} LLMs can support defensive analysis, log triage, malware explanation, vulnerability management, and incident response, but they can also lower the barrier to abuse \citep{deng2023pentestgpt}. The main deployment question is not whether the model can discuss security, but what tools, logs, credentials, and exploit-generation workflows it can access. Strong role-based access control, policy gating, controlled tool environments, and complete audit trails are essential.

\textbf{Software engineering.} Coding agents face a distinctive form of indirect prompt injection because repository content is both task data and a potential instruction carrier. Malicious instructions can appear in comments, documentation, tests, issue text, dependency metadata, build logs, or generated tool output. Secure design should treat repository content as untrusted unless explicitly trusted by the developer. Risk controls include sandboxed execution, dependency review, restricted file writes, approval before commits, and separation between code suggestion and code execution.

\textbf{Enterprise knowledge management.} Internal assistants often combine private documents, access control, \rag{}, and persistent memory. The main risk is that retrieval may bypass document permissions or mix information across users, teams, or tenants. Access control should be enforced before retrieval, not only after generation. Retrieved chunks should carry provenance and sensitivity metadata through the full pipeline.

\textbf{Robotics and physical systems.} The safety impact increases sharply when language plans control physical actions. A wrong instruction that would be merely inconvenient in a chatbot can become dangerous when translated into motion, navigation, manipulation, or equipment control. Systems require simulation, constrained controllers, runtime monitors, emergency stops, and human approval for risky operations. Natural-language plans should not directly map to actuator commands without safety validation.

\subsection{Practical Risk-Review Questions}

Before deployment, a risk review should answer the following questions:
\begin{enumerate}[leftmargin=*]
\item \textbf{Authority:} What external authority does the model or agent have, and which actions are irreversible or high impact?
\item \textbf{Trust boundaries:} What untrusted content can enter the prompt, retrieval context, memory, tool outputs, or system instructions?
\item \textbf{Data exposure:} What private, regulated, proprietary, or security-sensitive data can be retrieved, generated, stored, or logged?
\item \textbf{Tool containment:} Are tools allowlisted, scoped, sandboxed, and mediated by structured schemas and policy checks?
\item \textbf{Human oversight:} Which actions require human confirmation, and does the user see a faithful preview of the proposed action?
\item \textbf{Evaluation:} Has the system been tested against direct prompt injection, indirect prompt injection, jailbreaks, retrieval poisoning, privacy probes, and tool-misuse scenarios?
\item \textbf{Observability:} Can the organization reconstruct model version, prompt context, retrieved sources, tool calls, policy decisions, and approvals after an incident?
\item \textbf{Recovery:} Is there a rollback plan for model updates, poisoned documents, unsafe memories, leaked credentials, or harmful tool actions?
\end{enumerate}

These questions often reveal that the highest-value controls are architectural rather than linguistic. Narrower privileges, clearer provenance, safer retrieval, deterministic tool boundaries, and better incident evidence usually reduce deployment risk more reliably than adding additional natural-language instructions to the prompt.

\section{Open Problems}
\label{sec:open-problems}

The preceding sections show that LLM vulnerabilities are not isolated defects of a model, but emergent properties of a lifecycle and application stack. A deployed system combines training data, alignment procedures, prompts, retrieval pipelines, memory stores, tools, human users, organizational policies, and software dependencies. This coupling creates several research challenges that cannot be solved by improving model refusal behavior alone. We organize the open problems into three groups: \emph{foundational security problems}, which concern the boundary between language and computation; \emph{system-design problems}, which arise when LLMs are connected to data and tools; and \emph{assurance and governance problems}, which determine whether deployed systems can be evaluated, monitored, and repaired.

\subsection{Compositional Security Guarantees}

Current defenses rarely compose. A prompt-injection detector, a \rag{} filter, a safety classifier, and a tool schema may each perform well in isolation, yet fail when an attacker chains them across a long interaction. For example, a malicious document may first influence retrieval, then bias summarization, then trigger a tool call, and finally hide evidence in a generated report. This type of cross-layer failure is difficult to capture with single-turn attack success rates.

The central open problem is how to provide security guarantees for an application whose planner is nondeterministic, whose inputs include adversarial natural language, and whose intermediate states are partially hidden. Formal methods, information-flow control, capability systems, typed tool interfaces, and runtime monitors are promising directions, but they remain difficult to integrate with probabilistic language models. Future work should move from evaluating isolated defenses to evaluating \emph{defense compositions}. Useful questions include: whether a system preserves confidentiality under arbitrary retrieved text, whether an agent can invoke only authorized tools, and whether safety policies remain invariant under summarization, translation, compression, or multi-step planning.

\subsection{Instruction--Data Separation}

Indirect prompt injection exploits the absence of a reliable boundary between instructions and data. In conventional secure systems, code and data are separated by design; in many LLM applications, this boundary is blurred because both user commands and untrusted documents are represented as natural language in the same context window. Delimiters, role labels, and natural-language warnings are useful engineering practices, but they do not create a hard security boundary.

A major research direction is the design of stronger instruction--data separation mechanisms. Possible approaches include typed context regions, non-executable data blocks, provenance-aware attention mechanisms, structured intermediate representations, and external policy engines that decide which content may authorize an action. Another direction is to train or adapt models to respect context labels such as \texttt{instruction}, \texttt{untrusted document}, \texttt{tool output}, \texttt{memory}, and \texttt{policy}. The challenge is to make these labels semantically effective rather than merely decorative. A mature solution would allow an LLM to read untrusted content for information while preventing that content from changing goals, permissions, or tool policies.

\subsection{Provenance-Aware RAG and Long-Term Memory}

\rag{} and memory systems introduce persistent, externalized context. They improve factual grounding and personalization, but they also create new attack surfaces: poisoned documents, stale knowledge, cross-user leakage, embedding inversion, unauthorized retrieval, and memory manipulation. Current vector databases often optimize semantic similarity while carrying limited trust metadata. As a result, retrieved content may be relevant but unsafe, outdated, unauthorized, or adversarial.

Future \rag{} systems should preserve provenance across ingestion, chunking, embedding, indexing, retrieval, reranking, summarization, and answer generation. Open problems include embedding-level access control, poisoning-resistant retrieval, source conflict handling, freshness-aware generation, and verifiable deletion of memories. Memory systems require particular care because they can silently influence future sessions. Users and auditors should be able to inspect what is stored, why it was stored, when it expires, which future tasks may use it, and how it can be deleted. The long-term goal is not simply more accurate retrieval, but accountable retrieval with traceable authority.

\subsection{Secure Agency and Tool Delegation}

LLM agents require a theory of delegated authority. A user may ask an agent to plan a trip, summarize email, search a repository, or refactor code. The agent should infer helpful steps, but it should not invent new authority. This distinction is difficult because many useful tasks require planning, tool use, and interpretation of ambiguous intent.

Secure agency requires dynamic scoping of permissions to the user's current goal. Tool access should be narrow, temporary, auditable, and revocable. High-impact actions should require confirmation, and the confirmation should describe the actual external effect rather than merely restating the model's plan. For example, `send this email to these recipients with this attachment'' is a stronger confirmation than `continue?'' Evaluation should measure not only whether agents complete tasks, but also whether they avoid unsafe actions under malicious observations, poisoned tool outputs, or conflicting instructions. Future benchmarks should include realistic tool permissions, partial failures, deceptive documents, and multi-step attacks.

\subsection{Privacy-Preserving Adaptation}

Enterprises increasingly adapt LLMs using private documents, tickets, code, emails, logs, and user interactions. This creates privacy risks across both parametric and non-parametric components of the system. Fine-tuning may memorize sensitive examples; embeddings may leak semantic information; retrieval may cross tenant boundaries; logs may store private prompts; and model outputs may reveal information about training or retrieved data.

Differential privacy, federated learning, secure enclaves, synthetic data, redaction, and retrieval isolation provide partial solutions \citep{dwork2006differential,abadi2016deep,mcmahan2017communication}. However, the open challenge is to preserve utility while providing measurable guarantees for the whole application stack. A privacy guarantee for model parameters alone is insufficient if the retrieval store, prompt logs, cache, or memory layer remains exposed. Future work should develop end-to-end privacy accounting for LLM applications, including model adaptation, embedding generation, retrieval, logging, and human feedback collection.

\subsection{Robustness Under Distribution Shift and System Evolution}

LLM applications evolve continuously. Models are updated, prompts are revised, retrieval corpora change, tools are added, user behavior shifts, and attackers adapt. A defense that works at deployment time may degrade after a model update or after new documents enter the retrieval index. This creates a form of security distribution shift.

Open problems include safety regression testing, update-aware red teaming, continuous evaluation, and automatic detection of behavior drift. Model updates should be treated as security-relevant events, especially when the application has access to tools or private data. Similarly, retrieval-index updates should be tested for poisoning and permission errors. A mature deployment pipeline should evaluate not only model quality, but also whether existing safety, privacy, and tool-use guarantees still hold after each change.

\subsection{Evaluation Realism and Reproducibility}

Many current benchmarks underrepresent adaptive, long-horizon, stateful attacks. Real attackers can observe failures, revise prompts, poison future retrieval, exploit tool outputs, wait for maintenance windows, and chain small weaknesses across components. A single-turn jailbreak benchmark is therefore insufficient for evaluating deployed LLM systems.

Future benchmarks should include multi-session attacks, realistic permissions, benign hard cases, cost and latency constraints, multi-agent communication, retrieval poisoning, tool misuse, and adaptive attackers who know the defense. Evaluation should report not only attack success rate, but also false-positive rate, utility loss, cost, latency, user friction, and failure recoverability. Reproducibility is another open challenge because many studies depend on closed models, changing safety policies, and non-deterministic decoding. Shared benchmark harnesses, versioned prompts, fixed evaluation protocols, and transparent reporting checklists are needed for cumulative progress.

\subsection{Multimodal and Cross-Channel Attacks}

As LLM systems become multimodal, the attack surface expands beyond text. Instructions can be embedded in images, screenshots, audio, video frames, documents, tables, code, diagrams, or user-interface elements. A model may extract text from an image, summarize a PDF, interpret a webpage, or act on a screenshot. Each of these channels can carry untrusted instructions.

The open problem is to define security policies that survive cross-modal translation. A malicious instruction hidden in an image should not gain authority merely because it was converted into text by an OCR module or vision-language model. Similarly, a table cell, figure caption, metadata field, or audio transcript should carry provenance and trust labels after conversion. Future research should study multimodal prompt injection, cross-modal provenance, safe document parsing, and security-aware multimodal representation learning.

\subsection{Human Factors, Usability, and Over-Reliance}

LLM security is also a human factors problem. Users may over-trust fluent outputs, ignore warnings, approve unsafe tool calls, paste sensitive information, or misunderstand the system's authority. Conversely, overly strict defenses may create alert fatigue, reduce usefulness, or encourage users to bypass safeguards.

Open problems include designing effective confirmations, calibrated uncertainty displays, user-facing provenance, and warnings that are specific enough to change behavior. A useful confirmation should reveal the concrete consequence of an action, the data sources used, and the uncertainty or risk involved. More broadly, LLM systems should be designed so that users can understand what the system knows, what it can do, what it is not allowed to do, and when human expertise is required.

\subsection{Incident Response, Auditing, and Governance}

LLM incidents may involve prompts, retrieved documents, memories, model versions, tool calls, credentials, users, or third-party services. Organizations need runbooks for identifying affected sessions, revoking leaked credentials, quarantining poisoned documents, rolling back model versions, deleting unsafe memories, notifying users, and updating evaluation suites. Unlike traditional software incidents, the root cause may be a combination of natural-language input, model behavior, retrieval state, and tool execution.

Governance should specify who approves new tools, who owns safety regressions, who reviews model updates, how prompts and policies are versioned, and how incidents are documented. Auditing mechanisms should preserve enough evidence to reconstruct security-relevant events without retaining unnecessary sensitive content. Future work should develop standard incident taxonomies, evidence schemas, and post-incident evaluation methods for LLM applications.

\section{Discussion}

LLM vulnerabilities are often described as mysterious properties of neural networks, but many practical failures arise from familiar systems problems: weak input validation, confused-deputy behavior, excessive privilege, insecure supply chains, insufficient logging, and missing incident response. What is new is the interface: natural language can simultaneously express user intent, untrusted evidence, executable instructions, and social manipulation. This makes LLM systems especially vulnerable when developers rely on model judgment to enforce boundaries that should be enforced by architecture.

A lifecycle and application-stack view also helps avoid two common extremes. The first extreme is model fatalism: assuming LLMs are inherently unsecurable because prompt injection cannot be perfectly solved at the model level. The second is model optimism: assuming that better alignment or a stronger system prompt will solve application security. The realistic position is layered. Models should become more robust, but applications must still constrain authority, preserve provenance, verify outputs, and monitor behavior.

The most important design shift is to treat LLMs as components inside security-critical systems, not as the security boundary themselves. The model can summarize, reason, draft, classify, and propose. It should not be the only component deciding whether untrusted content can override policy, whether a tool may execute, whether private data may leave a boundary, or whether a memory should persist.

\section{Comparison with Existing Surveys and Frameworks}

\begin{table*}[h!]
\centering
\small
\caption{Positioning of this survey relative to representative surveys and practitioner frameworks. Checkmarks indicate primary coverage; partial coverage indicates that the topic is discussed but not used as an organizing axis.}
\label{tab:survey-comparison}
\begin{tabularx}{\textwidth}{L{3.2cm} c c c c c c c Y}
\toprule
\textbf{Work / framework} & \rotatebox{90}{Lifecycle} & \rotatebox{90}{RAG} & \rotatebox{90}{Agents} & \rotatebox{90}{Privacy} & \rotatebox{90}{Supply chain} & \rotatebox{90}{Defenses} & \rotatebox{90}{Evaluation} & \textbf{Main distinction} \\
\midrule
LLM security/privacy surveys \citep{yao2024llmsecurity,xu2025surveyattacks} & partial & partial & partial & $\checkmark$ & partial & $\checkmark$ & partial & broad attack catalogues and privacy/security synthesis \\
LLM trustworthiness/alignment surveys \citep{liu2023trustworthy,naveed2023comprehensive} & partial & partial & partial & partial & & $\checkmark$ & $\checkmark$ & emphasize model capabilities, alignment, and evaluation \\
Agent-security surveys \citep{ferrag2025agentthreats,ling2026secureagents,chu2026systematicagents} & $\checkmark$ & $\checkmark$ & $\checkmark$ & partial & partial & $\checkmark$ & $\checkmark$ & focus on agentic workflows and protocols \\
OWASP LLM Top 10 \citep{owasp2025llm} & partial & $\checkmark$ & $\checkmark$ & $\checkmark$ & $\checkmark$ & $\checkmark$ & & practitioner-oriented risk taxonomy \\
NIST AML taxonomy \citep{vassilev2025adversarial} & $\checkmark$ & partial & partial & $\checkmark$ & partial & $\checkmark$ & partial & general adversarial ML terminology and lifecycle stages \\
This survey & $\checkmark$ & $\checkmark$ & $\checkmark$ & $\checkmark$ & $\checkmark$ & $\checkmark$ & $\checkmark$ & unifies lifecycle stage, application stack, security objective, attacker capability, and defense layer \\
\bottomrule
\end{tabularx}
\end{table*}

Table~\ref{tab:survey-comparison} clarifies novelty. We do not claim that each vulnerability is new. Instead, our contribution is a systems organization that maps attacks and defenses to the lifecycle and application stack. This framing is useful for practitioners because mitigation responsibility often belongs to different teams: data engineers, model trainers, safety teams, MLOps engineers, application developers, security teams, compliance officers, and human operators.

\section{Conclusion}

This survey reframed LLM vulnerabilities through a lifecycle and application-stack taxonomy. We organized attacks across data collection, pretraining, post-training alignment, packaging and supply chain, retrieval and memory, prompting and inference, tool/agent execution, and deployment/maintenance. This view shows that many risks are not isolated model failures but cross-boundary failures involving data provenance, instruction hierarchy, delegated authority, persistent state, and operational controls. We synthesized defenses into a defense-in-depth architecture that combines deterministic controls, model-level robustness, retrieval and memory governance, tool containment, monitoring, red teaming, privacy engineering, and incident response. The central lesson is that trustworthy LLM systems require both safer models and safer systems. Future work should move from isolated jailbreak demonstrations toward compositional security, provenance-preserving RAG, secure agency, realistic stateful benchmarks, and operationally grounded governance.

\bibliographystyle{abbrvnat}
\bibliography{reference}

\end{document}